\documentclass[twocolumn,9pt]{article} % here we use the article class, rather than elsarticle

\usepackage[square,numbers,sort&compress,comma]{natbib}

\usepackage{amsmath}
\usepackage{amssymb}
\usepackage{caption}
\usepackage{graphicx}
\usepackage{latexsym}
\usepackage{times}
\usepackage[pagewise]{lineno}
\usepackage{hyperref}
\usepackage{placeins}
\usepackage{dblfloatfix} 
\usepackage{afterpage}
\usepackage{url}

\topmargin - 12pt % might need to be set to 0pt for some installations
\renewenvironment{abstract}%
              {% - begin definition
               \small% - select font
               {\bfseries \abstractname}% - select font
               \par% - end a paragraph (skip \parsep)
               \vspace{10pt}% - add vertical space
              }% - complete definition

\renewcommand\abstractname{Abstract}

\newcommand{\nomenclature}% - name of command
              [1]% - number of arguments
              {% - begin definition
               \bgroup% - begin a local group
               \flushleft% - turn on flushleft option
               \small\bf% - select font
               #1% - insert title text
               \par% - end a paragraph (skip \parsep)
               \egroup% - terminate local group
              }% - complete definition

\renewcommand{\section}% - name of command
              [1]% - number of arguments
              {% - begin definition
               \bgroup% - begin a local group
               \flushleft% - turn on flushleft option
               \small\bf% - select font
               \refstepcounter{section}% - increment counter
               \arabic{section}. #1% - insert title text
               \par% - end a paragraph (skip \parsep)
               \egroup% - terminate local group
              }% - complete definition

\renewcommand{\subsection}% - name of command
              [1]% - number of arguments
              {% - begin definition
               \bgroup% - begin a local group
               \flushleft% - turn on flushleft option
               \small\em% - select font
               \refstepcounter{subsection}% - increment counter
               \arabic{section}.% - insert title text
               \arabic{subsection}. #1% - insert title text
               \par% - end a paragraph (skip \parsep)
               \egroup% - terminate local group
              }% - complete definition

\renewcommand{\subsubsection}% - name of command
              [1]% - number of arguments
              {% - begin definition
               \bgroup% - begin a local group
               \flushleft% - turn on flushleft option
               \small\em% - select font
               \refstepcounter{subsubsection}% - increment counter
               \arabic{section}.% - insert title text
               \arabic{subsection}.% - insert title text
               \arabic{subsubsection}. #1% - insert title text
               \par% - end a paragraph (skip \parsep)
               \egroup% - terminate local group
              }% - complete definition

  \newcommand{\acknowledgement}% - name of command
              [1]% - number of arguments
              {% - begin definition
               \bgroup% - begin a local group
               \flushleft% - turn on flushleft option
               \small\bf% - select font
               #1% - insert title text
               \par% - end a paragraph (skip \parsep)
               \egroup% - terminate local group
              }% - complete definition

  \newcommand{\sectionbib}% - name of command
              [1]% - number of arguments
              {% - begin definition
               \bgroup% - begin a local group
               \flushleft% - turn on flushleft option
               \small\bf% - select font
               #1% - insert title text
               \par% - end a paragraph (skip \parsep)
               \egroup% - terminate local group
              }% - complete definition

\begin{document}

% -------------------------------------------------------------------- %
% -------------------------------------------------------------------- %
% -------------------------------------------------------------------- %

% -------------------------------------------------------------------- %

\small
\baselineskip 10pt

% -------------------------------------------------------------------- %
% -------------------------------------------------------------------- %
% -------------------------------------------------------------------- %
\setcounter{page}{1}
% -------------------------------------------------------------------- %
\title{\LARGE \bf Tensor Train Representation of High-Dimensional Unsteady Flamelet Manifolds}

\author{\large Sinan Demir$^{a,*}$, Pierson Guthrey$^{b}$, Jason Burmark$^{b}$, Matthew Blomquist$^{c}$,\\
Brian T. Bojko$^{d}$, Ryan F. Johnson$^{d}$\\[10pt]
{\footnotesize \em $^a$Advanced Propulsion and Power Department, Argonne National Laboratory, Lemont, IL 60439, USA}\\[-5pt]
{\footnotesize \em $^b$Lawrence Livermore National Laboratory, Livermore, CA 94550, USA}\\[-5pt]
{\footnotesize \em $^c$Department of Applied Mathematics, University of California, Merced, California 95343, USA}\\[-5pt]
{\footnotesize \em $^d$Laboratories for Computational Physics and Fluid Dynamics, Naval Research Laboratory, Washington, DC 20375-5344, USA}
}

\date{}  %%% Leave as is, do not add date;

% -------------------------------------------------------------------- %
% -------------------------------------------------------------------- %
% -------------------------------------------------------------------- %
\twocolumn[\begin{@twocolumnfalse}
\maketitle
\rule{\textwidth}{0.5pt}
\vspace{-5pt}

\begin{abstract} % 100 to 300 words.
This study, for the first time, investigates the use of tensor trains (TTs) to represent high-dimensional unsteady flamelet progress variable (UFPV) manifolds in chemically reacting computational fluid dynamics (CFD). The UFPV framework captures the thermochemical state of reacting flows using a reduced set of parameters and pre-computed manifolds, avoiding the need to transport all species or solve large stiff reaction systems. High-dimensional manifolds enhance accuracy by resolving coupled thermochemical effects critical in high-speed reacting flows but impose substantial memory demands. Here, a five-dimensional UFPV manifold is constructed and stored in the TT format to address this limitation. Several chemical mechanisms and table sizes are examined to evaluate TT compression performance and accuracy. The TT representation achieves significant memory reduction while preserving manifold fidelity and combustion behavior. A one-dimensional reacting-flow case using the discontinuous Galerkin (DG)-based JENRE\textsuperscript{\textregistered} Multiphysics Framework confirms that TT-compressed manifolds are interchangeable with standard UFPV tables. In addition to memory reduction, benchmark tests show that TT-based manifold sampling can achieve up to $2.4\times$ speedup relative to dense tensor evaluation. Although demonstrated for the UFPV combustion model, the proposed TT framework is broadly applicable to other tabulation-based combustion methodologies and provides a scalable alternative to machine learning (ML)-based approaches for representing high-dimensional combustion manifolds.
\end{abstract}

\vspace{10pt}

{\bf Novelty and significance statement}

\vspace{10pt}

This work presents a novel tensor train (TT) framework for representing and compressing high-dimensional unsteady flamelet progress variable (UFPV) combustion manifolds with detailed chemical mechanisms. The proposed approach enables efficient storage and evaluation of tabulated chemistry while preserving thermo-chemical fidelity and providing explicit \textit{a priori} error control through user-defined tolerances, in contrast to machine-learning-based approaches that mainly rely on training accuracy. This is significant because high-dimensional flamelet manifolds are increasingly required to capture coupled thermo-chemical effects in high-speed reacting-flow simulations but are often limited by prohibitive memory requirements. By enabling compact and scalable manifold representations that are well suited for parallel execution on modern computing architectures, the TT framework facilitates the practical use of detailed chemistry tabulation in large-scale CFD simulations, offers a physics-consistent alternative to machine learning (ML)-based tabulation strategies, and can be readily applied to other tabulated combustion modeling approaches.

\vspace{5pt}
\parbox{1.0\textwidth}{\footnotesize {\em Keywords:Tensor trains; Unsteady flamelets; Reacting flows; Manifold compression; Curse of dimensionality} }

\rule{\textwidth}{0.5pt}
*Corresponding author.
\vspace{5pt}
\end{@twocolumnfalse}] 

% \linenumbers
\section{Introduction\label{sec:introduction}} \addvspace{10pt}

The finite-rate chemistry (FRC) formulation with detailed chemical mechanisms provides high-fidelity predictions of combustion by directly solving the advection, diffusion, and chemical source terms for each species. However, the integration of large, stiff systems of ordinary differential equations (ODEs) across widely separated chemical time scales introduces severe numerical stiffness and requires substantial computational resources, particularly in high-speed reacting flow simulations. Moreover, baseline FRC formulations do not inherently account for turbulence-chemistry interactions (TCI), which need to be incorporated through additional modeling, commonly via transported moments and presumed or transported PDF approaches ~\citep{POPE1985119}. 

Flamelet-based combustion models offer a promising alternative for computationally affordable, scale-resolved simulations of high-speed reacting flows \citep{GONZALEZJUEZ201726, Drozda2020,SAGHAFIAN20152163}, as they can potentially incorporate detailed chemical mechanisms and TCI effects while significantly reducing computational cost. Among these, the unsteady flamelet progress variable (UFPV) approach ~\citep{SUN2022111841} is particularly effective for scramjet applications \citep{Dem23}, where accurate modeling of pressure gradients, shock-flame interactions, and unsteady chemical kinetics is critical. UFPV reduces the reacting flow state to a small number of transported scalars by precomputing one-dimensional unsteady flamelets that describe the local thermochemical state as functions of mixture fraction, progress variable, and other controlling parameters like scalar dissipation rate. During CFD simulations, thermochemical quantities such as species mass fractions and progress variable source terms are efficiently retrieved from flamelet tables, enabling predictive yet tractable simulations of complex high-speed combustion configurations. Importantly, the unsteady flamelet formulation allows auto-ignition and extinction processes to be captured naturally, without requiring ad hoc hotspot initialization of the reactive kernel.

However, accurate representation of the coupled flow-chemistry physics in such configurations often necessitates higher-dimensional flamelet manifolds, which include additional parameters such as pressure to capture compression-expansion effects and enthalpy to properly account for thermal energy variations induced by wall heat transfer. While these extensions substantially improve model fidelity, they also increase table dimensionality, leading to higher memory demands and interpolation costs, an issue commonly referred to as the \emph{curse of dimensionality}. To address this, several studies have employed artificial neural networks (ANNs) to replace or compress large flamelet tables~\citep{Owo20,Owo21,Dem23,CHANDRASEKHAR2026114635}. The Grouped Multi-Target ANN (GMT-ANN)~\citep{Owo20} clusters correlated species for efficient mapping, while the Mixture of Experts (MoE)~\citep{Owo21} partitions the manifold using nonlinear gating functions. More recently, another ML-based approach by \citep{CHANDRASEKHAR2026114635} investigated a hybrid surrogate framework that combines neural networks with kernel regression to directly model the progress variable source term while clustering correlated species for efficient species mass fraction prediction. Overall, these approaches have demonstrated important memory savings with high accuracy in large-eddy simulations (LES) of hydrocarbon flames and engines. 

Despite these advances, training ML-based models to accurately represent the progress variable source term across different fuels, particularly for chemically sensitive fuels, remains challenging due to the strong nonlinearity and orders-of-magnitude variation of the source term~\citep{Dem23}, and the predictive fidelity of ML-based models ultimately depends on the quality and diversity of the training data. Motivated by these limitations, the present work explores a physics-preserving alternative using TTs~\citep{Cic16} for efficient storage and reconstruction of high-dimensional UFPV manifolds. TT belong to the broader class of tensor network (TN) representations and correspond to a one-dimensional chain topology that enables compact low-rank approximation of high-dimensional tensors. TTs provide a compact, low-rank and compressed representation of multidimensional data, drastically reducing memory and computational requirements while preserving accuracy. To the authors’ knowledge, this study presents the first demonstration of TN-based compression applied to high-dimensional UFPV manifolds (e.g., five-dimensional tables) for high-speed reacting flow applications. TT compression is evaluated for tables generated from multiple mechanisms and sizes, including a GRI 3.0 ~\citep{Smith1999GRIMech30},  22-species hydrocarbon model ~\citep{LUO2012265} and the 9-species Burke's hydrogen mechanism~\citep{burke_h2}, showing excellent compression factors and negligible loss in predictive fidelity. In addition to memory reduction, the TT representation can also accelerate tabulated manifold evaluation relative to baseline dense table lookup.

\section{Methodology\label{sec:sections}} \addvspace{10pt}

In this section, we briefly introduce the UFPV combustion model and the TT approach used to efficiently represent high-dimensional manifolds. The formulation of the UFPV tabulated manifold is first summarized, followed by the TT representation employed to compress and evaluate the tabulated data.

\subsection{Unsteady Flamelet Progress Variable Approach\label{subsec:subsection}} \addvspace{10pt}

\begin{figure}[h!]
\centering
\includegraphics[width=192pt,trim=3 3 3 3,clip]{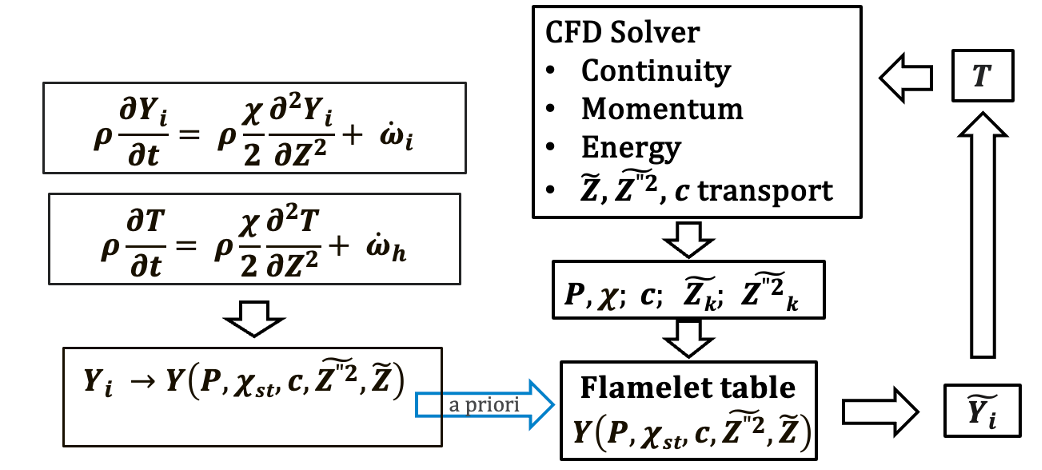}
\caption{\footnotesize 5D UFPV implementation in the JENRE\textsuperscript{\textregistered} Multiphysics Framework.}
\label{implementation}
\end{figure}

The UFPV approach implemented in the JENRE\textsuperscript{\textregistered} Multiphysics Framework (Fig.~\ref{implementation}), a high-order discontinuous Galerkin (DG) code operating on fully unstructured meshes, solves the unsteady flamelet equations using an in-house solver that leverages an analytical Jacobian and a sparse linear system solver under a prescribed set of flamelet boundary conditions~\citep{10.1115/1.4041281}. The resulting flamelet table stores species mass fractions and progress variable source terms over a range of independent variables. In the current UFPV formulation, the flamelet table is parameterized by five independent variables,
\begin{equation}
y_F = (Z, Z''^2, C, X_{st}, P),
\end{equation}
where transport equations are solved for the mixture fraction $Z$, its variance $Z''^2$, and the progress variable $C$.  The stoichiometric scalar dissipation rate $\chi_{st}$ characterizes the local mixing rate at the stoichiometric surface and controls the strain and extinction behavior of the flame, while $P$ represents the local pressure to account for compressibility effects.Within the compressible flamelet framework, temperature is computed from the transported enthalpy and retrieved species mass fractions via a Newton--Raphson iteration. Species gradients are reconstructed on-the-fly using finite differences rather than direct differentiation or tabulation~\citep{Boj23}.

Recent analyses of the UFPV implementation within the JENRE\textsuperscript{\textregistered} Multiphysics Framework demonstrate approximately $2$--$3\times$ speedup on both CPU and GPU relative to baseline FRC formulations for cases employing a relatively small chemical mechanism, with larger gains expected as mechanism complexity increases. Further details on the implementation, formulation, GPU-based optimization, accuracy, and high-speed reacting flow applications are provided in~\citep{demir_ufpv2026}.

\subsection{Tensor Trains for UFPV Tables\label{subsec:subsection2}} \addvspace{10pt}

Tensor networks have been recently gaining popularity due to their ability to compress multidimensional representations of data~\citep{Oseledets2011, lee2016fundamental, Oseledets2012}, accelerate neural networks~\citep{novikov2015tensorizing, phan2020stable}, enable efficient numerical methods for solving challenging PDE systems~\citep{guo2023local, MANZINI2023, DANIS2025113891}, advance various aspects of quantum computing~\citep{berezutskii2025tensornetworksquantumcomputing}, and, more recently, accelerate reacting flow simulations~\citep{JUNG2025117758}.

%\begin{figure}[h!]
\begin{figure*}[h!]
\centering
\includegraphics[width=300pt]{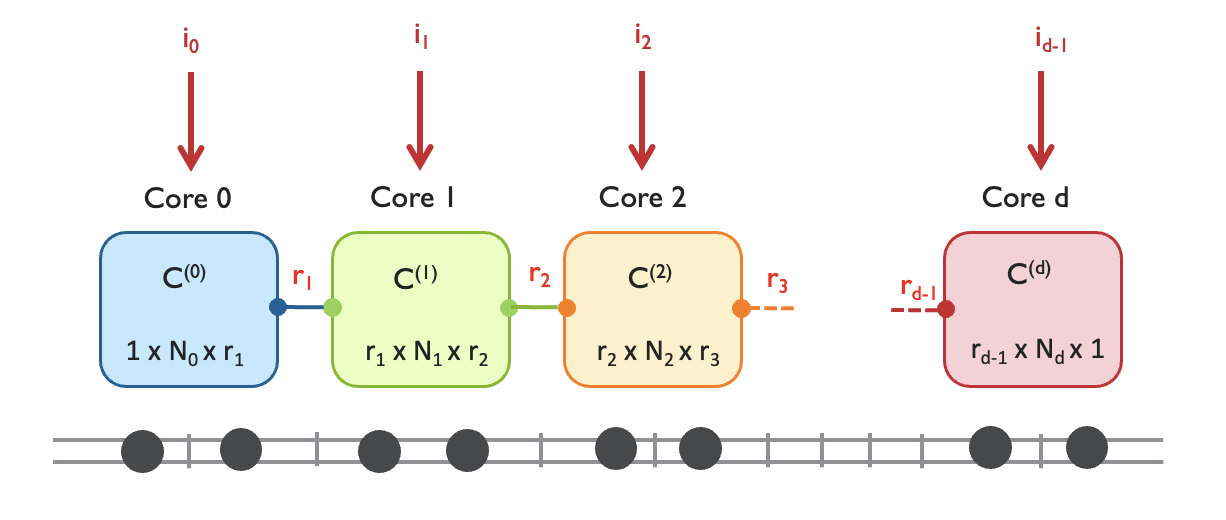}
\caption{\label{fig:tt_schematic} 
Schematic illustration of the tensor train (TT) decomposition of a $d$--dimensional table. 
Each TT core $C^{(k)} \in \mathbb{R}^{r_{k-1} \times N_k \times r_k}$ is a third--order tensor, where $N_k$ denotes the size of the $k$th table dimension and $r_k$ are the TT ranks connecting neighboring cores, with boundary ranks fixed as $r_0 = r_d = 1$. The arrows labeled $i_k$ indicate the physical indices of the original table, which select slices of each TT core during tensor evaluation, while the internal rank indices $r_k$ are contracted to reconstruct the tensor value.}
\end{figure*}

In this work we investigate the feasibility of representing the high-dimensional UFPV tables in the TT format. For example, given a pointwise defined function $f$ in three dimensions, we are using TTs to create an estimation $F$,
\begin{equation}
f_{ijk} \approx F_{ijk} = \sum\limits_{r_1}^{R_1} C_{r_0ir_1}^{(0)} \otimes \sum\limits_{r_2}^{R_2} C_{r_1jr_2}^{(1)} \otimes C_{r_2kr_3}^{(2)} ,
\end{equation}
where $N_i$ are the table extents,
$ijk$ are the table indices $i < N_0, j < N_1, k < N_2$,
$r_i < R_i$ for $i=0,1,2,3$ with $R_0=R_3=1$ are the TT ranks,
and $C^{(d)}_{\ell m n}$ for $d=0,1,2$ are third order tensors known as the TT cores. The ranks $R_i$ depend on a mix of the structure of the underlying data and a specified tolerance, see \cite{Oseledets2011}. The TT memory requirements are computable as
\begin{equation}
\text{memory, tensor train} = N_0R_1 + N_1R_1R_2 + N_2R_2,
\end{equation}
so if the ranks $R_i$ are small, the memory requirements of the TT table are much more favorable than the conventional lookup table.  We quantify this benefit as the compression factor, defined as
\begin{align}
\text{compression factor} 
&= \frac{\text{memory, full table}}{\text{memory, tensor train}} \notag\\[3pt]
&= \frac{N_0 N_1 N_2}{N_0 R_1 + N_1 R_1 R_2 + N_2 R_2}.
\end{align}

It is also important to quantify the estimation error introduced by the TT, which we simply compute as the relative error
\begin{equation}
\text{estimation error} = \frac{||f - F||_{\infty}}{|| f ||_{\infty}}.
\end{equation}

Figure~\ref{fig:tt_schematic} provides a generalized illustration of the TT representation used in this work. 
A $d$--dimensional lookup table is factorized into a sequence of low--rank third--order tensors (TT cores), each associated with one physical table dimension. 
During evaluation, the physical indices $i_k$ select slices of the corresponding TT cores, while the internal rank indices are contracted to reconstruct the table entry. 
This representation enables efficient storage and evaluation of high--dimensional UFPV manifolds while avoiding the exponential memory growth associated with full tabulation.

All TT decompositions and related tensor operations in this work are implemented using the $B\!\otimes\!B\vec{a}$ Tensor Network Library~\citep{BoBa}. Standalone $B\!\otimes\!B\vec{a}$ is used for a priori compression of high-dimensional UFPV manifolds and TT rank determination. The $B\!\otimes\!B\vec{a}$ is further interfaced with the JENRE\textsuperscript{\textregistered} Multiphysics Framework to enable in-solver TT-UFPV simulations and a posteriori tensor analyses. It provides efficient implementations of arbitrary-dimensional TT and Tucker decompositions suitable for high-performance CPU and GPU architectures. 

As an initial step, the objective of this study is to select a species-dependent tolerance such that the TT approximation error for each tabulated species remains within acceptable bounds. In the context of the UFPV framework, tabulated quantities are evaluated through multi-linear interpolation, which itself introduces a non-negligible numerical error. Accordingly, a natural and rigorous criterion for selecting the TT tolerance is to require that the approximation error be smaller than the interpolation error inherent to the tabulation procedure. This ensures that the TT compression does not introduce additional error beyond that already present in the baseline UFPV implementation. In this work, a simple estimate of the interpolation error is adopted and used to define the tolerance criterion, as described below.

\begin{equation}
\text{interpolation error} = \max\limits_{i < N_0, j < N_1, k< N_2 }
\left ( f^{max}_{ijk} - f^{min}_{ijk}
\right) ,  \label{interperror}
\end{equation}
where
\begin{equation}
f_{ijk}^{max} = \max\limits_{\hat i \in \{i,i+1\}, \hat j \in \{j,j+1\}, \hat k \in \{k,k+1\}  } f_{\hat i \hat j \hat k},
\end{equation}
and
\begin{equation}
f_{ijk}^{min} = \min\limits_{\hat i \in \{i,i+1\}, \hat j \in \{j,j+1\}, \hat k \in \{k,k+1\}  } f_{\hat i \hat j \hat k}.
\end{equation}

To illustrate the proposed tolerance selection strategy, Table~\ref{tab:tttable} shows compression factors and estimation errors as a function of the TT tolerance for an exemplar species, $CH_4$, using a representative two-dimensional manifold generated with the UCS mechanism~\cite{ZDV2023}. The interpolation error estimated using Eq.~(\ref{interperror}) is $0.0444987$. Accordingly, selecting a TT tolerance of $0.01$ yields an estimation error of the same order as the interpolation error while achieving a compression factor of $75\times$. Although this analysis must be repeated for each species and manifold, it demonstrates a principled criterion by which TT tolerances can be selected based on the inherent interpolation error of the tabulated manifold.

\begin{table}[h!]
\footnotesize
\centering
\caption{Compression factors and estimation errors for the CH$_4$ manifold as a function of tensor train (TT) tolerance.}
\begin{tabular}{ccc}
\hline
\text{Tolerance} &
\text{\begin{tabular}[c]{@{}c@{}}Compression \\ Factor\end{tabular}} &
\text{\begin{tabular}[c]{@{}c@{}}Estimation \\ Error\end{tabular}} \\
\hline
0.33    & 149 & 0.0722304 \\
0.10    & 149 & 0.0722304 \\
0.033   & 149 & 0.0722304 \\
0.01    & 75  & 0.0282988 \\
0.0033  & 50  & 0.0166388 \\
0.001   & 37  & 0.00775475 \\
0.00033 & 21  & 0.00384998 \\
0.0001  & 9   & 0.00164248 \\
\hline
\end{tabular}

\label{tab:tttable}
\end{table}

\section{Results and Discussion\label{sec:extext}} \addvspace{10pt}
\subsection{5D UFPV Table Compression: A Priori\label{subsec:subsection2}} \addvspace{10pt}
The following results examine TT compression performance for full high-dimensional (5D) UFPV manifolds, extending the illustrative single-species example to multi-species, CFD-relevant tables and quantifying the resulting compression–accuracy tradeoffs.

\begin{figure}[h!]
\centering
\includegraphics[width=192pt]{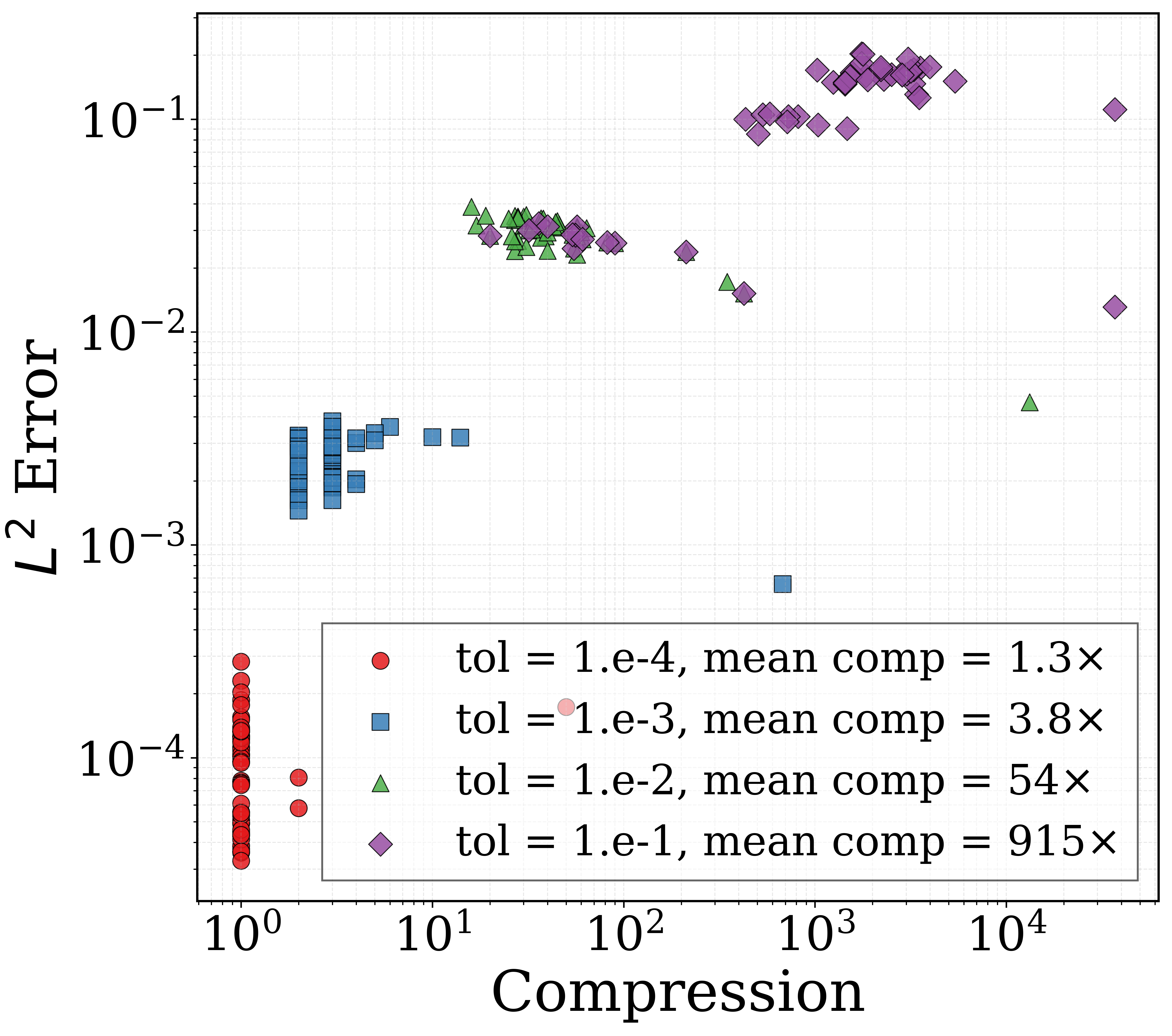}
\caption{\label{fig:scatter2} 
Relationship between tensor train (TT) compression and $L^2$ error for all species in the GRI 3.0 mechanism. Each color corresponds to a different tolerance, with the indicated mean compression factor.}
\end{figure}

Figure~\ref{fig:scatter2} shows the relationship between TT compression and relative $L^2$ error for all species in the GRI 3.0 ~\citep{Smith1999GRIMech30} mechanism. Each color represents a different tolerance level, with the corresponding mean compression factor indicated in the legend. As the tolerance is relaxed, higher compression factors are achieved at the expense of increased error. Conversely, tighter tolerances ($10^{-4}$–$10^{-3}$) maintain excellent accuracy but provide limited compression. Moderate tolerances ($10^{-3}$–$10^{-2}$) offer an optimal balance, yielding compression factors up to ${\sim}1000{\times}$ with minimal loss in fidelity.

%\begin{figure*}[ht!]
%\centering
%\vspace{-0.4 in}
%\includegraphics[width=300pt]{Figures/output_gri_scatter.png}
%\vspace{10 pt}
%\caption{\footnotesize
%Error associated with each individual species given a desired compression. 
%The colors represent different tolerances for acceptable error.}
%\label{fig:scatter}
%\end{figure*}

\begin{figure*}[ht!]
\centering
%\vspace{-0.4 in}
\includegraphics[width=300pt]{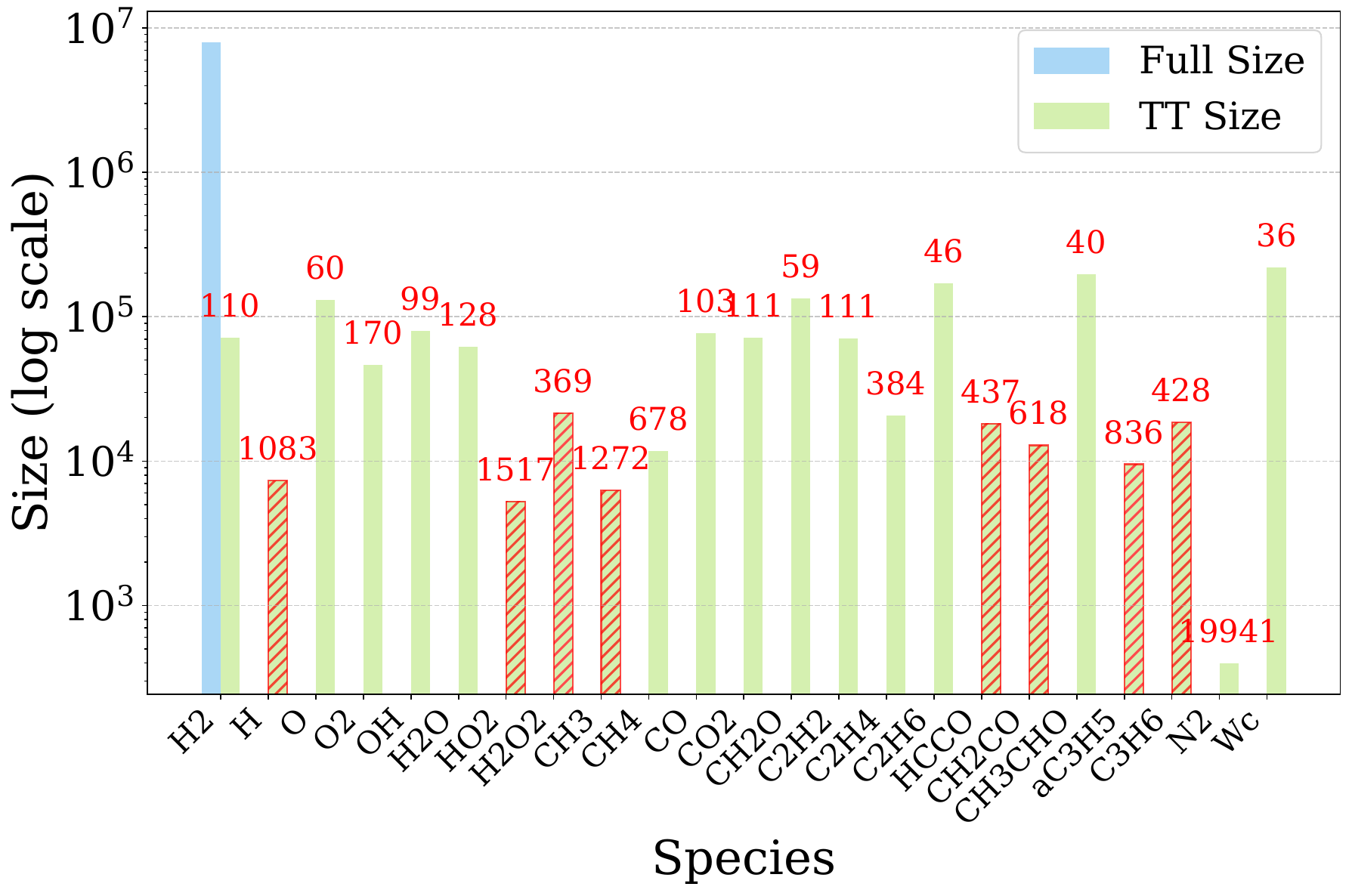}
%\vspace{5 pt}
\caption{\label{fig:histogram} 
Histogram of memory usage for a 22-species ethylene mechanism comparing a 5D UFPV table with its tensor train (TT) representation. The UFPV table requires $\sim$1.5~GB, whereas TT needs only $\sim$14.6~MB ($\sim$103$\times$ total compression). Red numbers denote per-species compression factors. Species bars with red diagonal hatch patterns denote minor species (tolerance 0.05), while solid-filled bars indicate major species (tolerance 0.01). }
\end{figure*}

To further assess the scalability of TT compression in practical CFD applications, Figure~\ref{fig:histogram} examines a 22-species ethylene mechanism previously used for scramjet simulations with the UFPV model~\cite{Dem23}. The UFPV table contains 22 species and is parameterized over five independent variables ($Z, Z''^2, C, X_{st}, P$). The stored layout corresponds to $(Y_i + w_c) \times Z \times Z''^2 \times C \times X_{st} \times P
= 23 \times 101 \times 12 \times 101 \times 13 \times 5$, where $23$ accounts for the 22 species mass fractions plus the progress-variable source term, and each entry is stored in double precision (8 bytes). In total, this table requires $\approx 1.5$~GB of memory, whereas the TT representation reduces the memory footprint to only 14.6~MB, yielding an overall compression factor of about 103$\times$. Compression efficiency varies across species: major species (H$_2$, O$_2$, OH, H$_2$O, CO, CO$_2$), constrained by a stricter tolerance of 0.01, achieve moderate compression factors, while minor species, treated with a looser tolerance of 0.05, exhibit much higher compressibility. This distinction is reflected in the Figure~\ref{fig:histogram}, where diagonal hatch patterns represent minor species. Radical intermediates such as HO$_2$, CH$_3$, and C$_3$H$_6$ display exceptionally high compression factors due to their localized distributions, whereas bulk species such as H$_2$O and CO$_2$ compress less aggressively. These results demonstrate that TT compression reduces overall storage requirements by two orders of magnitude while maintaining accuracy for critical species and enabling aggressive compression for less influential ones.

\begin{table*}[!ht]
\footnotesize
\centering
\caption{Memory usage and compression factors comparing UFPV vs. TTs for various table configurations ($Y_i + w_c \times Z \times Z''^2 \times C \times X_{st} \times P$), all reported in double precision(8 bytes).}

\begin{tabular}{cccc}
\hline
\text{Table Dimensions} & 
\text{\begin{tabular}[c]{@{}c@{}}UFPV \\ (MB)\end{tabular}} & 
\text{\begin{tabular}[c]{@{}c@{}}TT \\ (MB)\end{tabular}} & 
\text{\begin{tabular}[c]{@{}c@{}}Compression \\ Factor\end{tabular}} \\
\hline
$23 \times 101 \times 12 \times 101 \times 13 \times 5$ & 1500 & 14.6 & 103 \\
$23 \times 101 \times 12 \times 201 \times 13 \times 5$ & 2900 & 27.1 & 107 \\
$23 \times 201 \times 21 \times 201 \times 8 \times 5$ & 6200 & 15.0 & 417 \\
$23 \times 101 \times 12 \times 201 \times 16 \times 5$ & 12500 & 18.2 & 685 \\
$23 \times 101 \times 12 \times 201 \times 16 \times 7$ & 17500 & 21.4 & 816 \\
\hline
\end{tabular}
\label{tab:tt_compression}
\end{table*}

Table~\ref{tab:tt_compression} summarizes the memory requirements of UFPV and TT representations for several multi-dimensional table configurations. The dimensionality of each table grows with additional discretization points in mixture fraction, mixture fraction variance, scalar dissipation rate or pressure, leading to rapid increases in storage requirements for the baseline UFPV format. For instance, a table with dimensions $23 \times 101 \times 12 \times 201 \times 16 \times 7$ requires 17.5~GB in UFPV form, which is well beyond practical storage for large-scale LES/DNS applications. In contrast, the TT representation reduces the same case to only 21.4~MB, corresponding to a compression factor exceeding $800\times$. This trend highlights a key advantage of TT-based storage: while UFPV tables scale exponentially with the number of dimensions, TT representations scale linearly by exploiting redundancies in smooth, correlated manifolds of the thermo-chemical state space. Overall, the results show that TT compression transforms multi-gigabyte UFPV tables into compact, high-fidelity representations, making detailed, high-dimensional flamelet manifolds feasible for practical CFD applications.

\subsection{TT-UFPV CFD Simulation: A Posteriori\label{subsec:subsection2}} \addvspace{10pt}

\begin{figure}[h!]
  \centering
  \includegraphics[width=192pt]{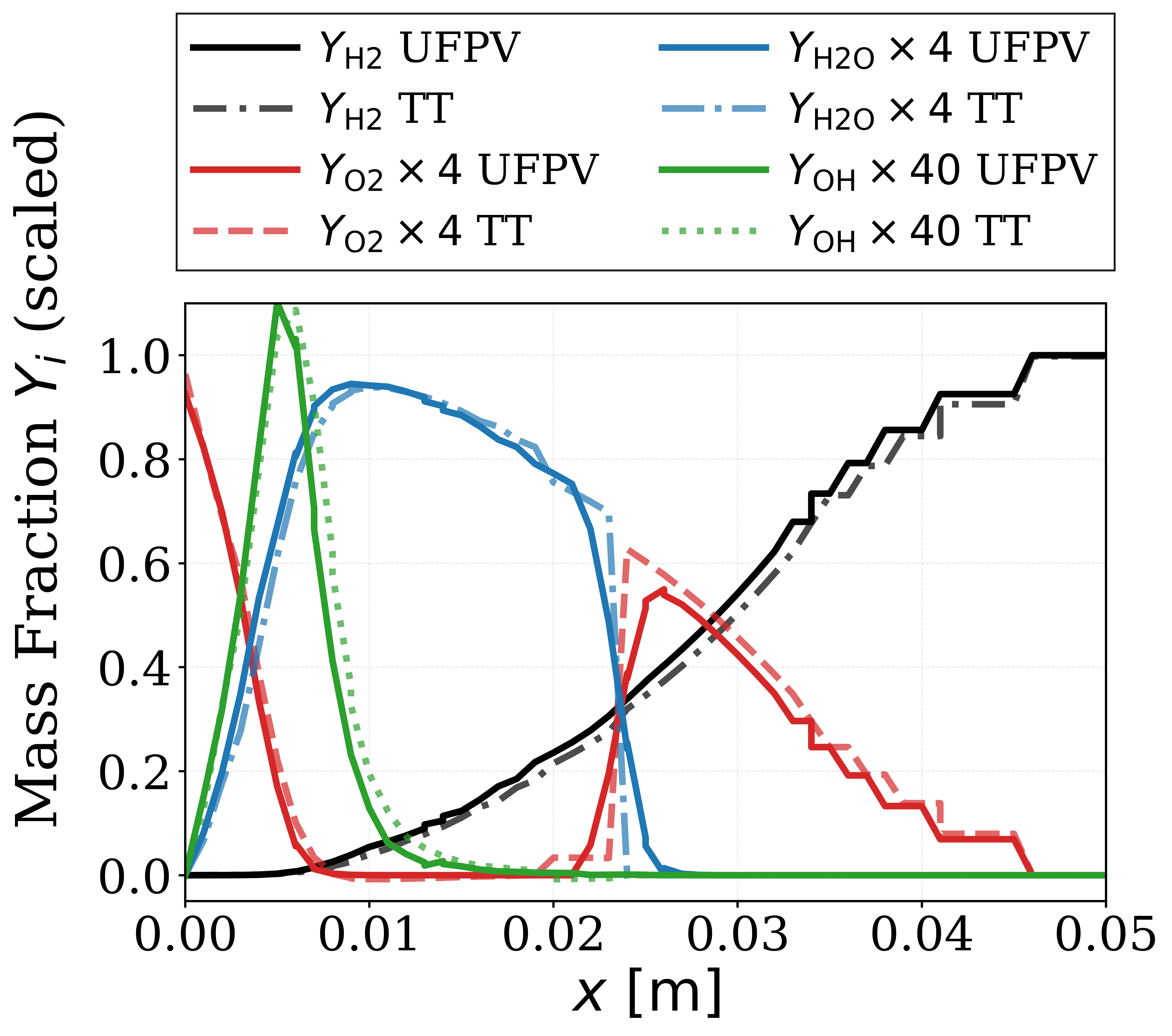}
  \caption{\footnotesize Comparison of UFPV and TT results for a 1D periodic flame for the selected species mass fractions
  ($Y_{H_2}$, $Y_{O_2}$, $Y_{H_2O}$, $Y_{OH}$) as functions of streamwise coordinate $x$.}
  \label{fig:single_species2}
\end{figure}

The recently implemented CPU and GPU-enabled UFPV-TT framework in the JENRE\textsuperscript{\textregistered} Multiphysics Framework has been validated through \textit{a posteriori} analyses of a one-dimensional periodic flame configuration, previously described and validated for UFPV simulations in~\cite{demir_ufpv2026}. Figure~\ref{fig:single_species2} compares line profiles of selected species mass fractions ($Y_{H_2}$, $Y_{O_2}$, $Y_{H_2O}$, $Y_{OH}$) between the baseline UFPV model and its TT-compressed counterpart. The five-dimensional UFPV manifold was generated using Burke’s hydrogen mechanism~\cite{burke_h2}, providing an additional test case with a different chemical mechanism and a hydrogen-based system known for strong fuel sensitivity. The TT representation using a tolerance of 0.01 preserves the major thermo-chemical structure and shows very close agreement with the baseline, with only minor deviations observed in regions of steep gradients, highlighting the balance between compression efficiency and fine-scale resolution. The demonstrated accuracy and robustness confirm that TT-compressed manifolds reproduce the baseline UFPV solution with negligible loss of fidelity, indicating that TT compression preserves the essential thermo-chemical structure required for CFD simulations.

\subsection{TT Computational Performance Evolution\label{subsec:subsection3}} \addvspace{10pt}

The computational performance of the TT representation was evaluated through a series of benchmark tests designed to isolate the cost of tabulated manifold evaluation. Dense tensor sampling was used as a baseline approach and compared with the TT-based evaluation for different sampling workloads. In this context, a sampling operation corresponds to evaluating the tabulated manifold for a given thermochemical state through multilinear interpolation of the table entries. The tabulated quantities are defined in a five-dimensional parameter space, and the number of sampling points is varied over several orders of magnitude to examine the scaling behavior of the TT evaluation procedure. Such sampling operations are frequently performed during CFD simulations that employ tabulated combustion manifolds.

Several TT core storage layouts were considered to assess the sensitivity of performance to the ordering of tensor indices in memory. Although these layouts correspond to mathematically identical TT decompositions, their organization in memory affects the access pattern during TT contraction and therefore influences the overall sampling cost. In addition to core storage layouts, different sampling patterns were also considered. Among these, the most demanding configuration corresponds to randomly distributed sampling points in the parameter space. This case minimizes spatial locality and therefore represents a worst-case access pattern for tabulated manifold evaluation.

The performance results for this configuration are shown in Fig.~\ref{fig:tt_performance}. Despite the unfavorable access pattern, the TT representation remains competitive with the dense tensor baseline and becomes increasingly advantageous as the number of sampling points grows. For the largest sampling counts considered, speedups exceeding a factor of two are observed, reaching approximately $2.4\times$ relative to dense tensor evaluation. This improvement arises from the reduced computational complexity of the TT representation, which replaces access to a full high-dimensional tensor with a sequence of low-rank contractions whose cost scales with the TT ranks rather than with the full tensor size. 

The numerical accuracy of the TT sampling was also verified for the benchmark tests by comparing the TT results against the dense tensor reference solution. The maximum difference between the two approaches was evaluated using the infinity norm with a tolerance of $10^{-9}$. In all tested cases, the observed errors were on the order of $10^{-12}$, confirming that the TT implementation reproduces the multilinear interpolation results to machine precision. This verification ensures that the measured performance differences arise from the computational efficiency of the TT evaluation rather than numerical discrepancies in the interpolation procedure.

  \label{fig:tt_performance}
\begin{figure}[h!]
  \centering
  \includegraphics[width=192pt]{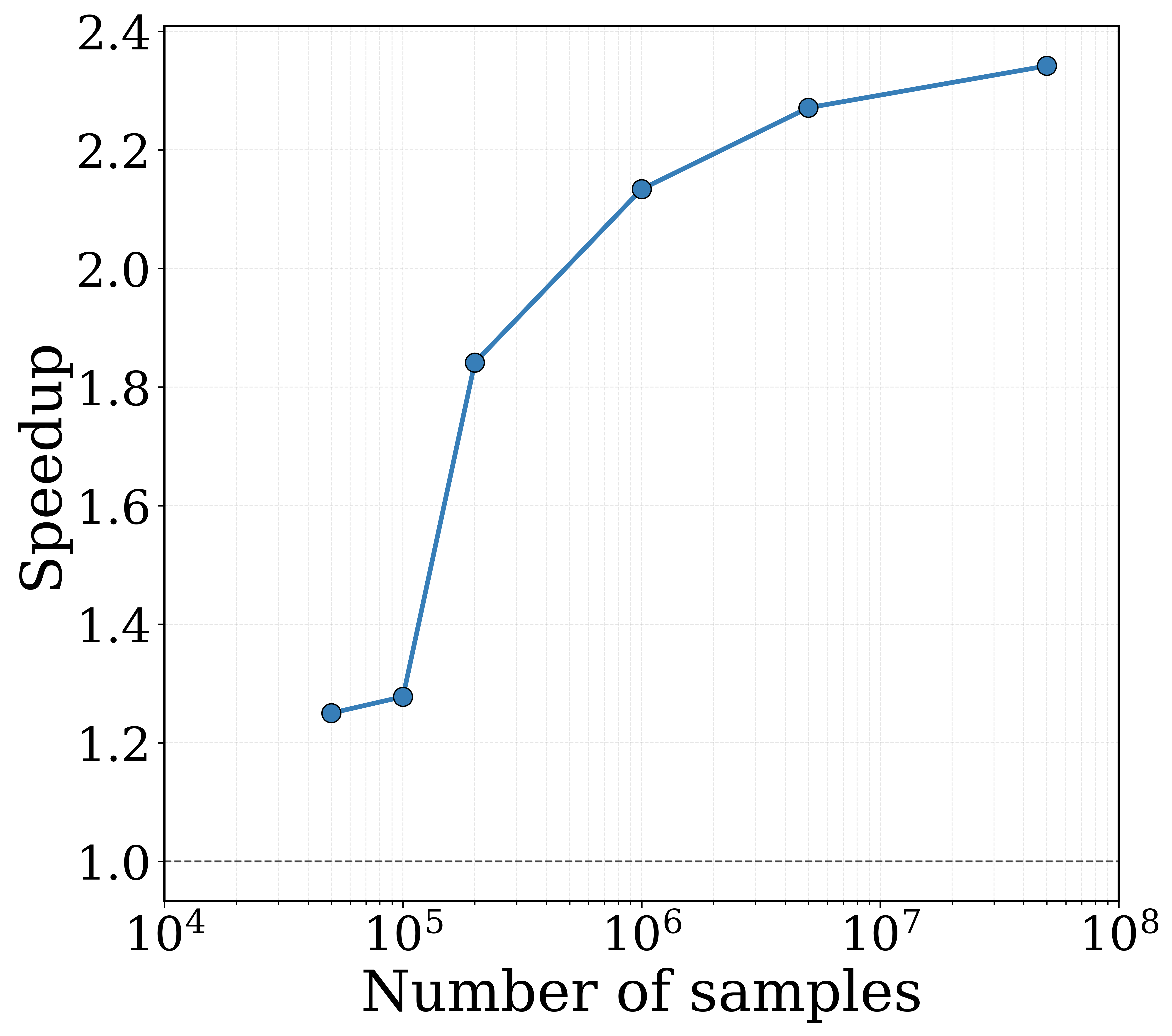}
  \caption{\footnotesize Performance comparison between dense tensor sampling and tensor train (TT) sampling for a five-dimensional UFPV manifold under randomly distributed sampling points. TT provides approximately $2.4\times$ speed-up. }
  \label{fig:tt_performance}
\end{figure}

\section{Conclusion\label{sec:figtabeqn}} \addvspace{10pt}

This study demonstrates the feasibility and effectiveness of TT decomposition for representing and compressing high-dimensional UFPV manifolds used in reacting-flow simulations. The TT representation mitigates the  \emph{curse of dimensionality} by replacing the exponential complexity growth of dense tensors with a low-rank factorization whose computational complexity scales approximately linearly with manifold dimension for moderate TT ranks. Through \textit{a priori} analysis, TT-based representations achieved orders-of-magnitude memory reduction while preserving the thermo-chemical accuracy of baseline flamelet data. Subsequent \textit{a posteriori} validation using one-dimensional periodic flame simulations in the JENRE\textsuperscript{\textregistered} Multiphysics Framework confirmed that TT-compressed manifolds accurately reproduce baseline UFPV predictions with negligible loss of fidelity. In addition to the memory reduction benefits, benchmark tests show that TT-based manifold sampling can also improve computational efficiency. For the five-dimensional manifolds considered in this study, TT sampling achieved speedups of approximately $2.4\times$ relative to baseline dense tensor evaluation on modern computing architectures. Unlike ML-based models, TT offers explicit \textit{a priori} error control through user-defined tolerances, providing predictable accuracy before deployment. The proposed TT framework is general and can be readily applied to other tabulated combustion models. These results establish TT compression as a scalable, physics-preserving alternative to data-driven manifold reduction methods and a key enabler for practical three-dimensional high-speed combustion simulations with detailed chemistry.

\acknowledgement{CRediT authorship contribution statement} \addvspace{10pt}

{\bf Sinan Demir}: Conceptualization, Methodology, Software, Investigation, Formal analysis, Visualization, Writing – original draft, Funding acquisition. 
{\bf Pierson Guthrey}: Conceptualization, Methodology, Software, Formal analysis, Writing – review and editing. 
{\bf Jason Burmark}: Software. 
{\bf Matthew Blomquist}: Formal analysis. 
{\bf Brian T. Bojko}: Software. 
{\bf Ryan F. Johnson}: Conceptualization, Methodology, Software, Formal analysis, Project administration, Writing – review and editing.

\acknowledgement{Declaration of competing interest} \addvspace{10pt}

The authors declare that they have no known competing financial interests or personal relationships that could have appeared to influence the work reported in this paper.

\acknowledgement{Acknowledgments} \addvspace{10pt}

The authors acknowledge support from Dr. Eric Marineau of the Hypersonic Aerothermodynamics, High-Speed Propulsion, and Materials Program, Office of Naval Research (ONR), Code 35. This manuscript was prepared in part by UChicago Argonne, LLC, Operator of Argonne National Laboratory (Argonne). Argonne, a U.S. Department of Energy Office of Science laboratory, is operated under Contract DE-AC02-06CH11357. Argonne National Laboratory’s work was supported by the Office of Naval Research (ONR). This work was also performed under the auspices of the U.S. Department of Energy by Lawrence Livermore National Laboratory under Contract DE-AC52-07NA27344.

% -------------------------------------------------------------------- %
% -------------------------------------------------------------------- %
% -------------------------------------------------------------------- %
\footnotesize
\baselineskip 9pt

% -------------------------------------------------------------------- %
% -------------------------------------------------------------------- %
% -------------------------------------------------------------------- %
\clearpage
\thispagestyle{empty}
\bibliographystyle{proci}
\bibliography{orgPROCI_LaTeX}

% -------------------------------------------------------------------- %
% -------------------------------------------------------------------- %
% -------------------------------------------------------------------- %

\newpage

\small
\baselineskip 10pt

% -------------------------------------------------------------------- %
% -------------------------------------------------------------------- %
% -------------------------------------------------------------------- %

\end{document}